\documentclass[aps,pra,twocolumn,showpacs,preprintnumbers,amsmath,amssymb,superscriptaddress,nofootinbib]{revtex4}

\usepackage{amsmath,amssymb}
\usepackage[latin1]{inputenc}
\usepackage{dcolumn}
\usepackage{bm}

\usepackage{graphicx}
\usepackage{dcolumn}
\usepackage{bm}

\begin{document}


\title{Effects of detector efficiency mismatch on security of quantum cryptosystems}

\author{Vadim Makarov}
\email{makarov@vad1.com}
\affiliation{Department of Electronics and Telecommunications, Norwegian University of Science and Technology, NO-7491 Trondheim, Norway}
\affiliation{Radiophysics Department, St.~Petersburg State Polytechnic University, Politechnicheskaya street 29, 195251 St.~Petersburg, Russia}

\author{Andrey Anisimov}
\affiliation{Radiophysics Department, St.~Petersburg State Polytechnic University, Politechnicheskaya street 29, 195251 St.~Petersburg, Russia}

\author{Johannes Skaar}
\affiliation{Department of Electronics and Telecommunications, Norwegian University of Science and Technology, NO-7491 Trondheim, Norway}

\date{\today}

\begin{abstract}
We suggest a type of attack on quantum cryptosystems that exploits variations in detector efficiency as a function of a control parameter accessible to an eavesdropper. With gated single-photon detectors, this control parameter can be the timing of the incoming pulse. When the eavesdropper sends short pulses using the appropriate timing so that the two gated detectors in Bob's setup have different efficiencies, the security of quantum key distribution can be compromised. Specifically, we show for the Bennett-Brassard 1984 (BB84) protocol that if the efficiency mismatch between 0 and 1 detectors for some value of the control parameter gets large enough (roughly 15:1 or larger), Eve can construct a successful faked-states attack causing a quantum bit error rate lower than 11\%. We also derive a general security bound as a function of the detector sensitivity mismatch for the BB84 protocol. Experimental data for two different detectors are presented, and protection measures against this attack are discussed.
\end{abstract}

\maketitle

\section{Introduction}

Quantum cryptography enables secure communication between two parties Alice and Bob, given a quantum channel and an authentic public channel \cite{bennett_brassard,ekert,bennett1992,zgisin}. The security is guaranteed by the laws of quantum mechanics \cite{mayers98,mayers96,lo_chau,shor_preskill} rather than assumptions about the resources available to a potential adversary. Although the protocol for secret key distribution, quantum key distribution (QKD), can be proved secure in principle, in the real world the system is not perfect. Flaws in the source and/or detector may be exploited by an eavesdropper (commonly called Eve) to collect information about the key without being discovered. Intuitively, it seems clear that when the imperfections are sufficiently small, the QKD protocol may still be secure. The impact of several imperfections has been discussed previously, and corresponding security bounds have been established \cite{mayers96,koashi_preskill,inamori_lutkenhaus_mayers,gott_lo_lutk_presk}.

\begin{figure}
\includegraphics[width=4.7cm]{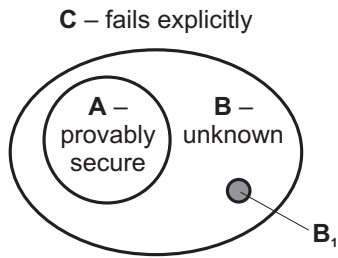}
\caption{\label{fig:security-spaces} Set of all possible input signals for a secure system.}
\end{figure}

Before we go on to consider a specific detector imperfection, let us discuss the place of our studies in the picture of security. For any system where security is required, the set of all possible input signals can be divided into three subsets (Fig.~\ref{fig:security-spaces}).
The subset A are the input signals for which the system is guaranteed to function normally (e.g., for a key distribution system, generate a secret key). The subset C are the input signals for which the system fails to perform the required function explicitly (e.g., fails to generate the secret key and alarms legitimate users about it). The subset B are input signals for which the system behaves in a way the developers are not quite sure about, thus potentially including subversions by a third party (e.g., generation of a key known to Eve while not raising an alarm). The last subset ideally should not exist and subsets A and C should ideally border one another, or at least the developers should be reasonably sure they do.

With classical digital systems requiring security, input data are binary strings, and the situation where the system is reasonably guaranteed to have empty subset B is achievable. For example, implementations of common cryptographic primitives are usually known to be reasonably secure. However, developers of protocols and applications with more complex functionality (e.g., most software for personal computers) often release them knowing that the subset B is likely nonempty; successful attacks would be found with time, and closed by applying patches on an {\em ad hoc} basis. The latter situation is clearly not acceptable for QKD.

The problem is that input data for Bob in a QKD system are not binary strings which are well defined and could be directly checked by an algorithm running on a classical computer. The input data for Bob are states of light that we, at the present level of technology, are having considerable difficulty detecting at all, and that have more degrees of freedom than binary data. This makes the important task of developing a complete security proof (and building a QKD system that fully corresponds to the model in the proof) intricate. We contribute to this effort by first showing that the subset B is still nonempty in the currently used model: some subset B$_1$ of input light states exists that results in a compromise of security, or that has merely not been considered before. Then, we try to find ways to expand the subsets A and C to cover B$_1$ via both extending the model in the proof and suggesting modifications to QKD setups.

More concretely, we will consider a specific imperfection at the detector; a mismatch in detector timing that occurs in most practical implementations of QKD over optical fibers. Most of today's quantum cryptosystems operating in the 1300 and 1550~nm telecommunication windows use gated avalanche photodiodes (APDs) as single-photon detectors. The detector is sensitive to an incoming photon for a short time (a few nanoseconds) called the detection window, and has practically zero sensitivity outside the detection window. The systems operate in a pulsed mode, where the expected time of photon arrival is synchronized with the middle part of the detection window. The systems have at least two separate detection windows or two separate detectors at Bob's side (for 0 and 1 bit values). These detection windows, while both covering the time when the photon comes, are inevitably shifted relative to each other, due to finite manu\-facturing tolerances. The shift may arise due to small optical path length differences or wire length differences, as well as other imperfections and variations in the detector electronics. Although the detector sensitivities might seem well matched when characterized with Alice's pulses, there may exist rapidly varying differences at the edges that can only be resolved with extremely short pulses. 

Eve may exploit a detector timing mismatch by using a version of the so-called faked-states attack \cite{makarov}. A faked-states attack on a quantum cryptosystem is an intercept-resend attack where Eve does not try to reconstruct the original states, but instead generates (quantum mechanical or classical) light pulses that get detected by the legitimate parties in a way controlled by her while not setting off any alarms. In this case, she may adjust the timing of her states in order to change the sensitivity of the 0 detector relative to that of the 1 detector, and vice versa. By using very short pulses she may take advantage of any rapidly varying features in the detector sensitivity curves not visible to Alice and Bob.

The paper is organized as follows. In Sec.\ II we introduce the faked-states attack in the ``ideal'' case where either detector can be totally blinded on Eve's choice. This attack gives Eve full information about the key while Bob registers no increase in the quantum bit error rate (QBER). In Sec.\ III we derive efficiency figures of a practically possible intercept-resend attack in a more realistic situation with partial efficiency mismatch. Section IV contains a discussion of the security for any eavesdropping attempts. Measurements of detector sensitivity curves for two different detectors are presented in Sec.~V. Finally, we discuss protection measures against this attack and conclude the paper in Sec.~VI. Although the attack is exemplified using the Bennett-Brassard 1984 (BB84) protocol \cite{bennett_brassard}, other protocols that use four states in two bases may also be vulnerable.

\section{Total detector sensitivity mismatch}

\begin{figure}[b]
\includegraphics[height=4cm,width=8.3cm]{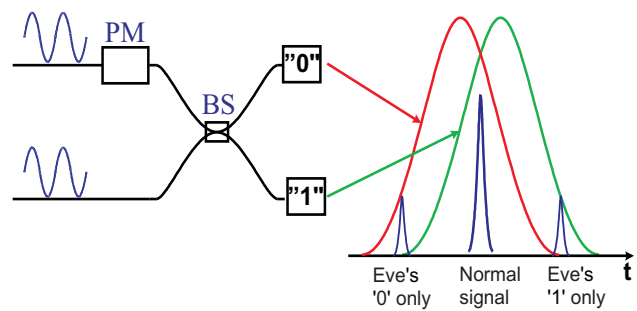}
\caption{\label{fig:det} Bob's part of the setup. Bob chooses the basis with the phase modulator (PM). The large detector efficiency mismatch is shown on the plot to the right.}
\end{figure}

To explain the attack, let us consider an ideal case when the detector sensitivity curves are significantly shifted in time relative to one another, so that time zones exist when one detector is completely blind while the other remains sensitive. Such a situation is depicted in Fig.~\ref{fig:det}.
The figure also shows the last part of the scheme with a Mach-Zehnder interferometer, a scheme example on which we will consider this attack.\footnote{Although a scheme with phase encoding is given as an example in Sec.~II, the attack and all obtained results equally apply to polarization encoding, owing to the formal isomorphism between the two encodings \cite{zgisin}.} During normal operation, Alice's pulse (denoted ``Normal signal'') is timed to the middle of the detector sensitivity curves, and both detectors are sensitive to it. Now if Eve mounts a faked-states attack, she cuts into the line and measures Alice's quantum states (choosing the basis randomly), and replaces them with faked states. She can construct faked states of pulses shifted in time to the sides of Bob's detector sensitivity curves, so that only one of the two detectors can fire in each case (the other one is blinded by timing). Thus she can set her bit value for Bob. Unlike the bit value, she has no direct control over which basis Bob applies with his phase modulator. However, Eve can make sure Bob never detects anything if he chooses a basis incompatible with Eve's measurement (which happens randomly in 50\% of the cases). To do this,  she sets the relative phase of the pulses in the two arms of the interferometer such that, if Bob chooses an incompatible basis and applies the corresponding phase shift to his phase modulator (PM), the interference outcome at the 50-50 coupler (BS) leads all light toward the detector that is blinded by timing. If, however, Bob chooses another basis (compatible with Eve's), the interference outcome at the coupler will be 50\%-50\% and the other detector will click. This trick works because, with today's components and transmission lines, Bob detects only a small fraction of the photons sent by Alice. The click at Bob's detector in the case of attack occurs with a reduced probability, but Eve can easily compensate by increasing the brightness of her faked states and thus keeping Bob's average detection rate the same as before mounting the attack. It is also easy to see that the bit statistics obtained by Bob is the same as that obtained in the absence of the attack. As you see, Eve now gets a complete copy of the key, and remains hidden.

The case of total detector sensitivity mismatch is not only convenient for explaining the principle of the attack, but can also occur in practice, as the experimental data later show. However, much more common and, indeed, {\em unavoidable} in reality would be the case when the detector sensitivities vary relative to each other in time but the ratio between them does not get very large. The implications of this property of detectors for security are analyzed in the rest of the paper.

\section{Partial detector sensitivity mismatch}

We will now consider the case when the sensitivity curves are slightly shifted, i.e., the detectors can only be partially blinded. For analysis in this section, we shall choose an eavesdropping strategy that is not necessarily optimal, but could clearly be implemented today. Let us simply adopt the intercept-resend strategy as described in the previous section for that.

Having chosen the strategy, let us consider all the possible basis and bit combinations during the attack. If we look at the relative phase of the pulses that Eve generates, we can note that, formally, she always chooses to resend to Bob the opposite bit value in the opposite basis compared to her detection. For example, if Eve detects a 0 in the $Z$ basis, she sends a 1 bit in the $X$ basis to Bob. She also chooses the timing so as to suppress 1 detection, i.e., a timing $t=t_0$ for which the ratio $\eta_1(t)/\eta_0(t)$ is small, where $\eta_0(t)$ and $\eta_1(t)$ denote the time-dependent detector efficiencies. The different events are shown in Table \ref{table:attack} for the special case where Alice sends a 0 in the $Z$ basis (the other three cases are symmetrical to this case). Initially, we assume that all states involved in the protocol and the attack are single-photon states. Later we will discuss the case where Alice and Eve use states with other photon statistics, e.g., faint laser pulses. Also, for now it is assumed that Bob's detectors have no dark counts (which is of course not true but we account for that later on). We assume that Eve's detectors and optical alignment are perfect, and that Eve generates faked states that match the optical alignment in Bob's setup perfectly. Based on the probabilities in the table we can now estimate the efficiency figures for this strategy in terms of the QBER and the mutual information between Eve and Alice, and Bob and Alice. 

\begin{table}
\newcommand\Ta{\rule{0pt}{2.5ex}}
\newcommand\Tb{\rule{0pt}{3.0ex}}
\caption{The intercept-resend attack when Alice sends a 0 in the $Z$ basis (as indicated in the first column). The second column contains the basis chosen by Eve and the measurement result; the third column shows the basis, bit, and timing as resent by Eve. In the next columns Bob's basis choice and measurement results are given. For the case with partial detector sensitivity mismatch, the probabilities for the different results are shown, given Eve's basis, bit value, and timing in addition to Bob's basis. Note that, for ease of discussion, the first two rows are repeated so that each row in the table occurs with probability 1/8.\smallskip
\label{table:attack}}
\begin{ruledtabular}
\begin{tabular}{c c c c c c c}
Alice\Ta & $\rightarrow$Eve & Eve$\rightarrow$ & Bob & \multicolumn{2}{ l }{Result, \  Probability} & Sifting\\[0.7ex]
\hline
$Z0$\Tb & $Z0$ & $X1t_0$ & $Z$&0, & $\frac{1}{2}\eta_0(t_0)$ & Keep\\[0.5ex]
 &  &  & &1, & $\frac{1}{2}\eta_1(t_0)$ & Keep\\[0.5ex]
 &  &  & &$-$, & $1-\frac{1}{2}\eta_0(t_0)-\frac{1}{2}\eta_1(t_0)$ & Lost\\[1.3ex]
$Z0$ & $Z0$ & $X1t_0$ & $X$&0, & $0$ & Discard\\[0.5ex]
 &  &  & &1, & $\eta_1(t_0)$ & Discard\\[0.5ex]
 &  &  & &$-$, & $1-\eta_1(t_0)$ & Lost\\[1.3ex]
$Z0$ & $Z0$ & $X1t_0$ & $Z$&0, & $\frac{1}{2}\eta_0(t_0)$ & Keep\\[0.5ex]
 &  &  & &1, & $\frac{1}{2}\eta_1(t_0)$ & Keep\\[0.5ex]
 &  &  & &$-$, & $1-\frac{1}{2}\eta_0(t_0)-\frac{1}{2}\eta_1(t_0)$ & Lost\\[1.3ex]
$Z0$ & $Z0$ & $X1t_0$ & $X$&0, & $0$ & Discard\\[0.5ex]
 &  &  & &1, & $\eta_1(t_0)$ & Discard\\[0.5ex]
 &  &  & &$-$, & $1-\eta_1(t_0)$ & Lost\\[1.3ex]
$Z0$ & $X0$ & $Z1t_0$ & $Z$&0, & $0$ & Keep\\[0.5ex]
 &  &  & &1, & $\eta_1(t_0)$ & Keep\\[0.5ex]
 &  &  & &$-$, & $1-\eta_1(t_0)$ & Lost\\[1.3ex]
$Z0$ & $X0$ & $Z1t_0$ & $X$&0, & $\frac{1}{2}\eta_0(t_0)$ & Discard\\[0.5ex]
 &  &  & &1, & $\frac{1}{2}\eta_1(t_0)$ & Discard\\[0.5ex]
 &  &  & &$-$, & $1-\frac{1}{2}\eta_0(t_0)-\frac{1}{2}\eta_1(t_0)$ & Lost\\[1.3ex]
$Z0$ & $X1$ & $Z0t_1$ & $Z$&0, & $\eta_0(t_1)$ & Keep\\[0.5ex]
 &  &  & &1, & $0$ & Keep\\[0.5ex]
 &  &  & &$-$, & $1-\eta_0(t_1)$ & Lost\\[1.3ex]
$Z0$ & $X1$ & $Z0t_1$ & $X$&0, & $\frac{1}{2}\eta_0(t_1)$ & Discard\\[0.5ex]
 &  &  & &1, & $\frac{1}{2}\eta_1(t_1)$ & Discard\\[0.5ex]
 &  &  & &$-$, & $1-\frac{1}{2}\eta_0(t_1)-\frac{1}{2}\eta_1(t_1)$ & Lost\\[1.7ex]
\end{tabular}
\end{ruledtabular}
\end{table}

We discard all cases where Alice and Bob have chosen incompatible bases. When Alice sends a 0 in the $Z$ basis, the probability that the qubit arrives at Bob is 
\begin{equation}
\label{ParrZ0}
P(\text{arrive}|A=Z0)=\frac{1}{4}[\eta_0(t_0)+\eta_0(t_1)+2\eta_1(t_0)].
\end{equation}
The probability of arrival averaged over Alice's four choices is found by symmetrization of this expression, yielding
\begin{equation}
\label{Parr}
P(\text{arrive})=\frac{1}{8}[\eta_0(t_0)+3\eta_0(t_1)+3\eta_1(t_0)+\eta_1(t_1)].
\end{equation}
Similarly, we find the QBER,
\begin{eqnarray}
\label{eq:QBER}
&& \text{(QBER)}=\frac{P(\text{error})}{P(\text{arrive})} \\
&& \ \ \ \ \ \ \ \ \ \ \  =\frac{2\eta_0(t_1)+2\eta_1(t_0)}{\eta_0(t_0)+3\eta_0(t_1)+3\eta_1(t_0)+\eta_1(t_1)}, \nonumber
\end{eqnarray}
where $P(\text{error})$ accounts for the cases when Bob detects a bit value different from what Alice has sent.

Having established the QBER, we will now compare Bob's and Eve's amount of relevant information \cite{shannon48}. Denoting the mutual information between Alice and Bob $H(A:B)$, and the mutual information between Alice and Eve $H(A:E)$, the security is guaranteed when $H(A:B)>H(A:E)$ \cite{csiszar}. This condition is sufficient and necessary for protocols with only one-way classical communications (no advantage distillation \cite{maurer93}; with advantage distillation it is not necessary). For intercept-resend attacks, it is clear that $A\rightarrow E\rightarrow B$ is a Markov chain. Hence, $H(A:B)\leq H(A:E)$, so Bob's key is generally not secure. Note that advantage distillation is not possible because intercept-resend attacks remove any entanglement between Alice's and Bob's qubit.

To analyze in more detail how this particular attack performs we will evaluate the mutual information between Alice and Eve $H(A:E)\equiv H(A)-H(A|E)$. After the basis has been revealed, $A$ takes only two possible values (0 and 1) while Eve's result is $Z0$, $Z1$, $X0$, or $X1$. We assume that Alice and Bob have used the $Z$ basis (by symmetry in the QKD protocol and the eavesdropping strategy we need only consider this basis choice). The entropy $H(A)$ is found from the probabilities $P(A)$, which, in turn, can be calculated from the arrival probabilities \eqref{ParrZ0} and \eqref{Parr}: 
\begin{subequations}
\label{PA}
\begin{align}
& P(A=0)=\frac{\eta_0(t_0)+\eta_0(t_1)+2\eta_1(t_0)}{\eta_0(t_0)+3\eta_0(t_1)+3\eta_1(t_0)+\eta_1(t_1)},\\
& P(A=1)=1-P(A=0).
\end{align}
\end{subequations}
To identify the conditional entropy $H(A|E)$, we need the conditional probabilities $P(E|A)$, and also $P(A|E)$ which can be found using Bayes' rule:
\begin{equation}
P(A|E)=\frac{P(A)}{P(E)}P(E|A).
\end{equation}
The conditional probabilities $P(E|A)$ are calculated using Table \ref{table:attack}:
\begin{subequations}
\label{PEA1}
\begin{align}
& P(E=Z0|A=0)=\frac{\eta_0(t_0)+\eta_1(t_0)}{\eta_0(t_0)+\eta_0(t_1)+2\eta_1(t_0)},\\
& P(E=Z1|A=0)=0,\\
& P(E=X0|A=0)=\frac{\eta_1(t_0)}{\eta_0(t_0)+\eta_0(t_1)+2\eta_1(t_0)},\\
& P(E=X1|A=0)=\frac{\eta_0(t_1)}{\eta_0(t_0)+\eta_0(t_1)+2\eta_1(t_0)}.
\end{align}
\end{subequations}
In the case $A=1$ we find the conditional probabilities directly from \eqref{PEA1} using the symmetry. The probabilities $P(E)$ are found using the relation
\begin{equation}
P(E)=\sum_a P(E|A=a)P(A=a),
\end{equation}
and the conditional entropy is
\begin{eqnarray}
&& H(A|E) \\
&& =-\sum_{e,a} P(A=a) P(E=e|A=a)\log P(A=a|E=e). \nonumber
\end{eqnarray}
After substitution of the probabilities above, the result is simple: $H(A|E)=\text{(QBER)}$, where the QBER is given by Eq.~\eqref{eq:QBER}. Hence,
\begin{equation}
H(A:E)=H(A)-\text{(QBER)}.
\end{equation}

The mutual information between Alice and Bob, $H(A:B)\equiv H(A)-H(A|B)$, is found by a similar procedure. After the basis has been revealed $A$ and also $B$ take only two values (0 and 1). The conditional probabilities $P(B|A)$ are
\begin{subequations}
\label{PBA}
\begin{align}
& P(B=0|A=0)=\frac{\eta_0(t_0)+\eta_0(t_1)}{\eta_0(t_0)+\eta_0(t_1)+2\eta_1(t_0)},\\
& P(B=1|A=0)=1-P(B=0|A=0),\\
& P(B=1|A=1)=\frac{\eta_1(t_1)+\eta_1(t_0)}{\eta_1(t_1)+\eta_1(t_0)+2\eta_0(t_1)},\\
& P(B=0|A=1)=1-P(B=1|A=1).
\end{align}
\end{subequations}

\begin{figure}
\includegraphics[height=5.7cm,width=7cm]{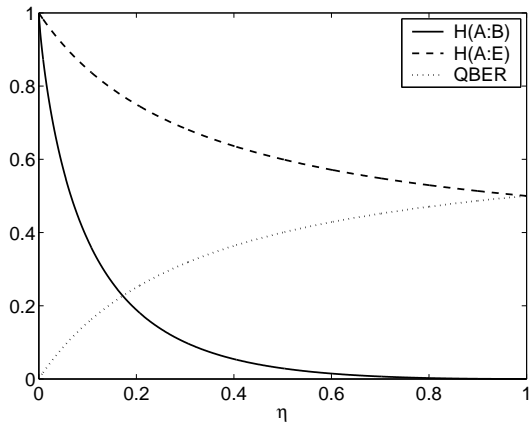}
\caption{\label{mutinf} The QBER, the mutual information between Alice and Bob, $H(A:B)$, and the mutual information between Alice and Eve, $H(A:E)$, as functions of the normalized efficiency of the blinded detector, $\eta$.}
\end{figure}

In the special case with symmetric detector efficiency curves, i.e., $\eta_0(t_0)=\eta_1(t_1)$ and $\eta_0(t_1)=\eta_1(t_0)$, we find $H(A:B)=1-h(\text{QBER})$ and $H(A:E)=1-\text{QBER}$, where $h$ is the binary Shannon entropy function $h(x) = - x\log_2 x - (1-x) \log_2 (1-x)$. Thus all quantities, the QBER, $H(A:B)$, and $H(A:E)$, depend only on one parameter; the normalized efficiency $\eta\equiv \eta_1(t_0)/\eta_0(t_0)$. The result is plotted in Fig. \ref{mutinf}.
As mentioned previously, it is apparent that Eve has always more mutual information with Alice than does Bob. For $\eta=1/3$ the difference $H(A:E)-H(A:B)$ reaches its maximum $h(1/3)-1/3\approx 0.58$ for a corresponding QBER of $1/3$. If Bob is not aware of his detector efficiency mismatch, he thinks that the key is secure when the QBER is less than 0.11 (symmetric protocols with one-way classical communications \cite{shor_preskill}). Thus Eve can compromise the security of the system if $\eta\leq 0.066$. The privacy amplification \cite{bennett95} Alice and Bob apply will not save them from this attack and will not produce a secret key because the mutual information between Alice and Eve is always greater than that between Alice and Bob.

In a real installation, Alice and Bob may expect the QBER to stay at some level below 0.11, which leaves Eve less room for the attack. Also in the practical scenario considered in this section, the contribution of dark counts in Bob's detectors to the total QBER is independent of other error sources and is beyond the control of Eve. Only the part of QBER {\em not} caused by dark counts in Bob's detectors can be used by Eve.

Let us consider any side effects this attack may produce that may divulge it. Although the attack may not give any alarm in terms of the QBER, it might be detected as a result of different measurement statistics at Bob's detector. From \eqref{PA} and \eqref{PBA} and their analogs for the case where Bob used the $X$ basis (incompatible basis), we observe that the measurement statistics has changed as a result of Eve's attack. However, the changes may be reduced or even eliminated by choosing suitable $t_0$ and $t_1$. (For example, the bit rates are equal in the symmetric situation analyzed above.) Similar skews in statistics may be produced in the absence of Eve's attack by random drifts and optical misalignments during operation, and may lie within what Bob normally expects.

So far we have assumed that Alice and Eve use single-photon states. Then Bob can detect the attack as a decreased bit rate, because $P(\text{arrive})$ usually would be less than the detection probability Bob has with no attack. Any reasonably well implemented Bob would monitor the bit rate and raise alarm if it drops significantly. To compensate for the reduced detection probability, Eve could increase the brightness of her pulses (several photons in each pulse, and possibly different photon statistics for the $t_0$ and $t_1$ pulses). However, this compensation might be possible to detect from the coincidence count rates at Bob's detectors. Alternatively, Eve could place her intercept unit and resend unit at two separate locations along the transmission line, thus winning the photons that would be lost in the line between these two locations. In the limit we have to assume she would place the intercept unit near Alice and the resend unit near Bob, getting the whole amount of normal loss in the line to cover for the reduction in detection probability caused by her attack.

If Alice uses faint laser pulses, the attack is still possible. However, now Eve must consider the basis-dependent coincidence count rates at Bob's detectors. If we grant Eve a future technology, namely, the ability to do photon number measurement, she would be able to retain the coincidence rates: Eve could measure the photon number first, and run the faked-states attack only on those pulses that contain one photon, using a single-photon source to generate faked states. Those of Alice's pulses that contain two or more photons can be passed undisturbed to Bob at the expense of a small part of the key becoming unavailable to Eve. Alternatively they can be eavesdropped on using the photon number splitting (PNS) attack \cite{felix,brassard_lutkenhaus,lutkenhaus}, provided a version of the PNS attack that does not alter coincidence counts could be constructed in this case \cite{lutkenhaus_2002_pns_keep_photon_statistics}.

Watching the rates and coincidence statistics for different bit-basis combinations is useful as a general precaution and should be built into the key distribution protocol. But it does not necessarily provide security against this attack.

\section{Security bound}

The intercept-resend attack described in the previous section is not necessarily the optimal attack. Alice and Bob want, of course, their protocol to be secure against any attack permitted by quantum mechanics. Note that Eve can exploit rapidly varying features in the detector sensitivity behavior even though she does not regenerate the pulses. She may perform a quantum nondemolition measurement of Bob's pulses to collapse them into much shorter ones, obtaining the associated timing information of the resulting pulse. As shown in the Appendix, this measurement will not disturb the degrees of freedom encoding Bob's qubit.

The following discussion of security will be based on the proofs by Lo and Chau \cite{lo_chau} and Shor and Preskill \cite{shor_preskill}. Here, Eve is allowed to do collective attacks and perform arbitrary quantum operations on each block of data. Alice and Bob use only one-way classical communications in the QKD protocol. Note that higher bit error rates can be tolerated if they use two-way classical communications \cite{gottesman_lo} (advantage distillation).
 
The critical point in the Lo-Chau and Shor-Preskill proofs is to bound the so-called bit and phase error rates. In the entanglement purification protocol used in the proof, this corresponds to bounding the fidelity of the Bell pairs received by Alice and Bob, and therefore the mutual information Eve has with their measurement results. In the QKD protocol, Alice and Bob measure the error rate by sampling a subset of the qubits randomly. Bob measures the qubits in two bases (chosen randomly for each qubit). The error rate as measured in the random sampling process is denoted the bit error rate; the error rate if Bob had chosen the opposite basis is denoted the phase error rate. In the case where Eve can control the detector efficiencies, we distinguish between the measured bit error rate (QBER) and the actual bit error rate. The measured bit error rate (QBER) is the error rate as measured by Bob, while the actual bit error rate is the error rate that Bob would measure if his detectors were perfect.

An analysis of several attacks where the eavesdropper has some information on the basis used by Bob is described by Gottesman {\em et al.}\ \cite{gott_lo_lutk_presk}. In the Trojan pony attack (Ref.\ \cite{gott_lo_lutk_presk}), the eavesdropper can control the efficiency of the detectors to create an asymmetry between the bit error rate (which is measured by Bob) and the phase error rate (which is not measured). In the optimal case (as seen from Eve's viewpoint) all errors that Eve eliminates are bit errors. Note that, in this case, the bit error rate as measured by Bob is the actual bit error rate since Eve does not control the two detector efficiencies separately (as opposed to the situation analyzed in this paper). Bob's problem is rather that he cannot measure bit and phase errors on the same qubit.

\begin{figure}
\includegraphics[width=8.1cm]{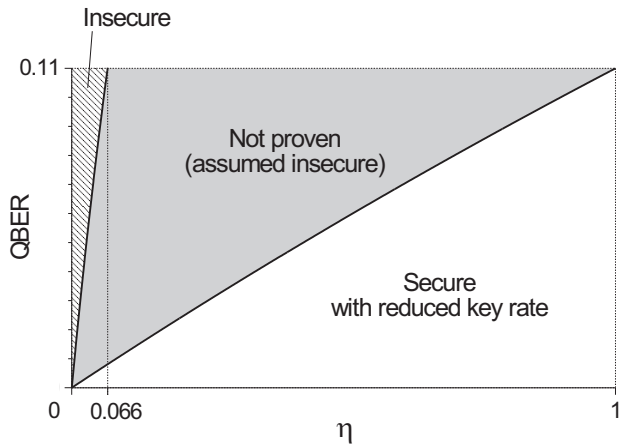}
\caption{\label{secchrt} Security state of a QKD system as a function of the normalized efficiency of the blinded detector $\eta$ and the measured QBER. In the ``Secure'' zone, the required amount of privacy amplification is larger than without considering this attack, being determined by $\delta$ given in Eq.~\eqref{securitybound}. In order to make this plot, we have allowed for some simplifications. The border between ``Not proven'' and ``Insecure'' zones is drawn assuming the special case of symmetric detector efficiency curves discussed in Sec.~III. The QBER for the ``Insecure'' zone is assumed to be without contribution from dark counts in Bob's detectors.}
\end{figure}

Now, consider the case relevant to the present paper, where Eve has no information on the basis used by Bob. Instead she can control the 0 and 1 detector efficiencies separately, by appropriate timing of the qubits. Since Eve does not know Bob's basis, the actual bit and phase error rates will be equal. However, since Eve can force the efficiencies of the two detectors to be different, the measured bit error rate will be different from the actual bit error rate. Therefore, Bob has to estimate the actual bit error rate from the measured bit error rate and {\em a~priori} knowledge of Eve's power (that is, he must characterize his detector sensitivity curves).

The available bit rate from the QKD after privacy amplification is \cite{shor_preskill}
\begin{equation}
R = 1 - 2 h(\delta),
\end{equation} 
where $\delta$ is the actual bit error rate and $h$ is the binary Shannon entropy function $h(x) = - x\log_2 x - (1-x) \log_2 (1-x)$. The actual bit error rate is related to the measured error rate and the detector efficiencies. The two detector efficiencies are denoted $\eta_0(t)$ and $\eta_1(t)$, and at a certain time $t$, they may be different. For example, take $\eta_0(t) > \eta_1(t)$. In a worst-case scenario, Eve minimizes the measured bit error rate (QBER) for a given $\delta$. Assuming a large number $N$ of qubits, $\delta N$ of them would be detected as errors if the detectors were perfect. For Bob's detectors, in the worst case this number is reduced to $\eta_1(t)\delta N$ provided Eve uses the timing $t$. At the same time, the number of qubits detected as correct bits is only reduced from $(1-\delta)N$ to $\eta_0(t)(1-\delta)N$. The associated QBER becomes $\eta_1(t)\delta/[\eta_1(t)\delta+\eta_0(t)(1-\delta)]$. Minimizing with respect to $t$, we obtain\footnote{Eve may certainly use several different $t$'s for different qubits. However, since ${\sum_i p_i}/{\sum_i q_i}\geq \min_i ({p_i}/{q_i})$ for any positive $p_i$ and $q_i$, the minimum QBER is still given by the minimum of $\eta_1(t)\delta/[\eta_1(t)\delta+\eta_0(t)(1-\delta)]$ and $\eta_0(t)\delta/[\eta_0(t)\delta+\eta_1(t)(1-\delta)]$ for all $t$.}
\begin{equation}
\label{securityboundQBER} 
\text{(QBER)} = \frac{\eta\delta}{1+\eta\delta-\delta},
\end{equation}
where
\begin{equation}
\label{etadef} 
\eta=\min\left\{\min_t \frac{\eta_1(t)}{\eta_0(t)}, \min_t \frac{\eta_0(t)}{\eta_1(t)}\right\}.
\end{equation}
In other words, the estimate for $\delta$, 
\begin{equation}
\label{securitybound} 
\delta = \frac{\text{(QBER)}}{\eta+(1-\eta)\text{(QBER)}},
\end{equation}
and not the QBER, should be used to determine the required amount of privacy amplification. The QKD protocol is secure provided $\delta<0.11$ [0.11 is the zero of $1 - 2 h(\delta)$], which means approximately that $\text{(QBER)}<0.11\eta$.

The bound above might be a little pessimistic: Eve needs at least a ``partial'' qubit measurement to decide which timing to use for the pulses going to Bob. This measurement must certainly be performed before Eve gets information on the basis used by Alice and Bob. The Shor-Preskill bound assumes that Eve may wait with her measurement until the basis choice is made public.

The security findings that have been made in the paper are summarized in Fig. \ref{secchrt}.

\section{Experimental data}

In this section we present measured detector sensitivity curves of two different single-photon detectors. Both devices under test were laboratory prototypes of detectors that were a part of or intended for use in quantum cryptography systems.

\subsection{Detector model~1}

\begin{figure}
\includegraphics[width=8.1cm]{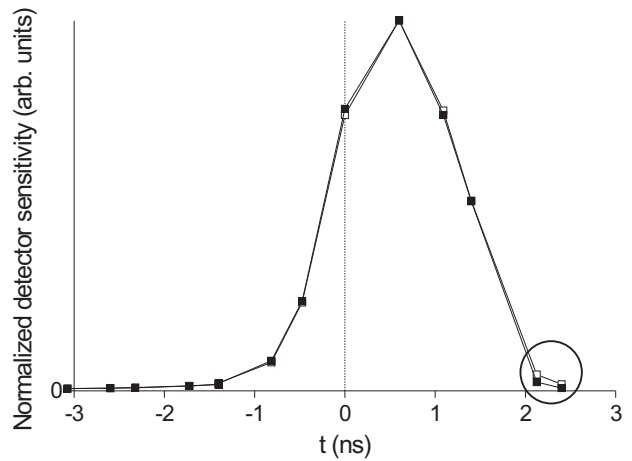}
\caption{\label{sc-tr} Detector model~1. Sensitivity curves for the 0 (open squares) and 1 (filled squares) time slots, at low mean number of photons at the APD ($\mu\ll1$). Dark counts were subtracted. The curves, originally of different height, were scaled so that their peak points coincide. $t$ is the relative time of arrival of the laser pulse at the APD; $t=0$ was the actual arrival time of Alice's pulse in the operational QKD setup before this measurement.}
\end{figure}

The first detector we tested was a time-multiplexed detector, i.e., a single detector registering 0 and 1 counts in different time slots. The light pulses corresponding to the 0 and 1 bit values were combined into a single fiber (one of the pulses was delayed in an optical delay line), and fed to the detector. The detector was gated at double the pulse rate, with 0 pulses coming in odd gates and 1 pulses coming in even gates. The model operated at 1310~nm and used a Soviet-made Ge APD (standard part number FD312L, developed by NPO Orion) cooled to 77~K. Gate pulses at the APD in this detector were made as narrow as practically possible, around 2~ns full width at half maximum (FWHM). The laser pulse in the test was 100~ps wide (FWHM) and was actually the same pulse normally used by Alice: we simply employed the entire QKD setup described in Ref.\ \cite{makarov_brylevski} to do the detector test, only changing the time delay of the laser pulse in order to measure the sensitivity curves. The measured curves are presented in Fig.~\ref{sc-tr}.

Since the same detector is used for 0 and 1 detections, we would expect the shapes of sensitivity curves to be highly identical. This is indeed the case. Also the curves have almost no time shift relative to one another, which means the fiber optic delay line in our setup was cut and spliced with good precision (from these data we can estimate the cutting inaccuracy to be less than $\pm25$~ps or $\pm5$~mm). Nevertheless the time range (encircled on the chart) where the laser pulse impinges the APD at the closing edge of the gate shows sensitivity mismatch $\eta\approx1/2$. It is possible the mismatch is actually larger than this, but we could not resolve it unless we used narrower laser pulses and did a more detailed measurement in this time range. The other side of the peak where the laser pulse impinges the APD before and at the opening edge of the gate shows no discernible sensitivity mismatch, because the APD sensitivity in this time range rises smoothly. This is consistent with the presence of a trailing tail in a typical APD time response \cite{lacaita_zappa,zappa_lacaita}.

The measured curves suggest that the practical attack described in Sec.~III would be impossible, but the general security bound \eqref{securitybound} would impose a significant penalty on the key rate and maximum allowed QBER. It is also clear that a better measurement with narrower laser pulse (no wider than few tens of picoseconds), smaller time increments, and extended time range would generally be desired for detector testing.

The precision with which the fiber delay line was cut in this setup was actually unnecessary for normal operation of the QKD. Should less care be taken in cutting the delay line, there would typically be larger mismatch at both sides of the curve. In the worst possible case one of the curves could end up shifted to the left by 1.1~ns, providing the same sensitivity for Alice's pulse as we have now while leaving sufficiently large mismatch at the sides for Eve to attempt the practical attack described in Sec.~III.

\subsection{Detector model~2}

The second detector we tested was a dual detector, consisting of two identical single-photon detectors registering 0 and 1 counts in parallel. This detector was one of the several different test prototypes developed at the Radiophysics Department at the St.~Petersburg State Polytechnic University. Each of the two detector channels had its own APD, gating, and detection electronics, while the thermoelectric cooler for the APDs, power supply, and external synchronization were shared. JDS Uniphase EPM239BA (former Epitaxx EPM239BA) single-mode fiber pigtailed APDs were used, cooled to $\approx-48~\textdegree$C. The APDs were gated at 100~kHz, with gate pulses having magnitude of 8~V and width of 3.5~ns (FWHM). The laser pulse in the test had wavelength of 1560~nm and was less than 200~ps wide (FWHM). The detector was set into a mode that would be suitable for its operation in a QKD system. The peak efficiencies in both channels were made to be roughly equal, by adjusting the bias voltage separately on each APD. The laser pulses impinged both APDs almost simultaneously; the remaining small difference in the optical paths, 9~mm or 45~ps between the channels, was later accounted for when plotting the charts so they represent the response to a laser pulse impinging both APDs at exactly the same time.

\begin{figure}
\includegraphics[width=8.1cm]{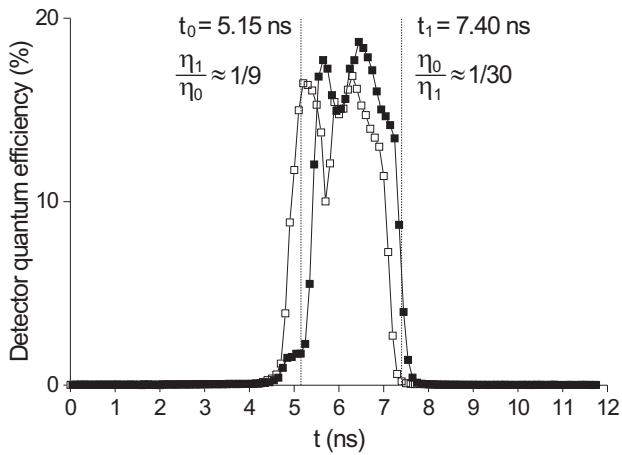}
\caption{\label{sc-sp05} Detector model~2. Sensitivity curves for the 0 (open squares) and 1 (filled squares) time slots, at mean number of photons at the APD $\mu=0.5$. Dark counts were subtracted.}
\end{figure}

With this detector, we tried to do a more thorough measurement than with the previous one. The sensitivity curves are shown in Fig.~\ref{sc-sp05}.
Although the curves overlap in a 1.6-ns-wide zone (well enough for use in QKD), there are significant mismatches at the sides. Using the time $t_1$ marked on the chart, and a $t_0$ where both detector efficiencies are small, Eq.~\eqref{eq:QBER} gives $\text{QBER}\approx 0.061$. This may give an impression that the attack described in Sec.~III is possible. However, any properly implemented Bob would raise an alarm if the 0 and 1 detection rates were significantly different. To achieve more similar detection rates, Eve can increase the brightness of her $t_0$ pulses and/or tune $t_0$. In the limit where the two detection rates are equal, she chooses the $t_0$ as marked on the chart to obtain the minimum QBER of 0.119. This means that the attack would be discovered (however, it is close to the threshold). Nevertheless, the QKD system with this detector will be rendered inoperative by the general security bound \eqref{securitybound}, which for $\eta=1/30$ allows a QBER of no more than 0.0036. Note that shifting the curves relative to one another never eliminates large sensitivity mismatch.

\begin{figure}
\includegraphics[width=8.1cm]{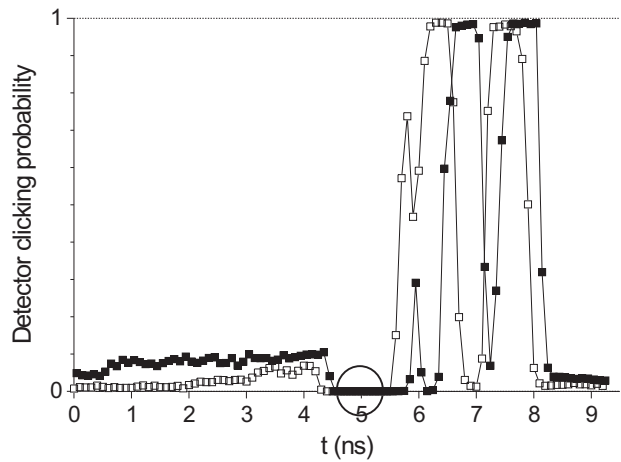}
\caption{\label{sc-sp500} Detector model~2. Sensitivity curves for the 0 (open squares) and 1 (filled squares) time slots, at mean number of photons at the APD $\mu=500$. In the encircled time range (4.65--5.30~ns) the clicking probability in both detectors measured exactly zero (0~counts registered per $>10^5$ gates). Unfortunately the time reference in this plot is not accurately matched with that in Fig. \ref{sc-sp05}, and the curves' features cannot be directly compared between the two figures.}
\end{figure}

In the measurement above, we could not see the quantum efficiency in the long tails, because it was masked by dark counts. It was therefore natural to repeat the measurement using three orders of magnitude brighter pulses. The expected result is complete saturation in the middle, and elevated, well-resolved tails. The result we obtained, however, was quite surprising (Fig.~\ref{sc-sp500}).
Although the measurement did resolve the tails (showing a significant mismatch around 1~ns), the detector performance in the middle part of the chart was erratic, with sensitivity plunging to zero where there should have been saturation. Using this behavior of the detector, Eve could likely run the attack in conditions close to the total sensitivity mismatch described in Sec.~II.

\begin{figure}
\includegraphics{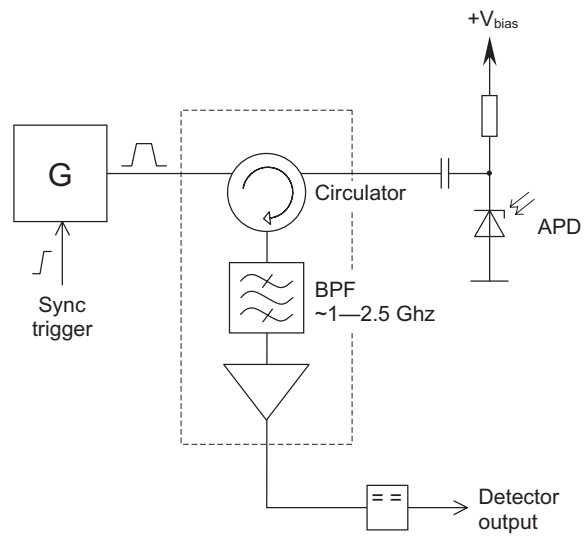}
\caption{\label{spb-det-diagr} Detector model~2. Equivalent diagram of a single channel. G is a single-shot generator that forms the gate pulse for the APD. BPF is an equivalent band-pass filter representing the frequency bandwidth of the tract for the reflected signal.}
\end{figure}

Forced to explain this detector behavior, we turned to the schematic of its electronics. The feature of this particular test prototype was that it used signal reflected from the APD, so that only one electrical waveguide had to be connected to each APD, thus reducing the thermal flow and easing cooling (Fig.~\ref{spb-det-diagr}).
To split off the reflected signal, a microstrip coupler was used, forming a circulator at frequencies above 1~GHz. The following amplifier had the bandwidth of ca.\ 2.5~GHz. Thus the whole tract for the reflected signal suppressed spectral components outside the 1--2.5~GHz band. There was no balancing circuit for spikes in the reflected signal that resulted from the gate front and back edges causing current through the APD capacitance, and also the spikes seeping into the reflected signal tract through other electrical imperfections. These unwanted spikes were partially suppressed spectrally: most of the spectrum of the spikes lay below 1~GHz, as the front and back edges of the gate pulse were less steep than the front edge of the avalanche signal. The comparator threshold was fine tuned to be lower than the avalanche signal, but higher than the parasitic signal at the output of the tract in the absence of avalanche. This all worked fine for avalanches caused by absorption of 1--2 photons, as Fig.~\ref{sc-sp05} illustrated. However, with avalanches caused by almost simultaneous absorption of hundreds of photons from every laser pulse, this spectral-selective circuit connected to a finely tuned comparator produced the gaps seen in Fig.~\ref{sc-sp500}.
The use of a spectral-selective circuit was a necessary condition for this abnormal behavior. The spectrum of the avalanche pulse was a function of two varying parameters: the pulse length and the shape of its front edge. The fraction of the avalanche pulse that passed through the spectral-selective tract to the comparator thus depended on these two parameters. Small changes in them due to the use of brighter light pulses resulted in the observed behavior of the output signal. Exact details of APD operation with brighter pulses, however, proved to be elusive to measure with the equipment we had.

Although we were able to eliminate the abnormal detector behavior with $\mu=500$ laser pulses by making adjustments in the electronics, this test prototype together with the idea of using reflected signal and/or spectral-selective detection tract had to be scrapped. It is simply too risky from the security standpoint to use detectors based on this or any other ``advanced'' approach in QKD systems, even if you test them well. More straightforward detection schemes have to be preferred.

\section{Discussion and Conclusion}

We have seen that when the detection of 0 and 1 bits can be blinded separately by timing, Eve can obtain full information about the key while she is hidden. In the case with only partial sensitivity mismatch, a similar attack is possible which will not provide alarm to Alice and Bob in terms of the QBER when the mismatch is sufficiently large. Although the specific intercept-resend attack given in Sec.\ III only works in certain conditions, more sophisticated attacks may exist which are able to exploit small sensitivity mismatches. Hence, to ensure secure QKD it is crucial to characterize Bob's detectors and specify maximum sensitivity mismatch. Based on this information, the worst-case estimate for $\delta$ given in \eqref{securitybound}, and not the QBER, should be used to determine the required amount of privacy amplification.

Specific measures aimed to specify and/or limit the sensitivity mismatch might be the following.

(1) Measure detector characteristics (especially sensitivity vs time) over a variety of input signals, including those well {\em beyond} the normal operating range. Use sufficiently short pulses so that all features of the sensitivity curves are captured. Employing a simple, straightforward detector circuitry can help lower the likelihood of hidden surprises, both discovered and undiscovered by testing.

(2) Introduce intentional random jitter in the detector synchronization to ``smear'' the curves and lower the mismatch.

(3) Implement active protection by checking timing of incoming pulses at Bob. This can be done through random shifting
of Bob's detection time window, by registering the time of avalanche onset within the window, or with additional detectors.

In the future it would be desirable to see if the general security bound, as implied by \eqref{securitybound}, can be narrowed. The security bound as it stays now is rather strict, and requires the amount of privacy amplification to be corrected in most practical quantum cryptosystems that use four-state protocols.

Not all QKD protocols are vulnerable to this attack. For example, the Bennett 1992 (B92) protocol \cite{bennett_b92orig,koashi_2004,muller_1997,z_merolla_1999} is not affected, because it uses just one detector for quantum states (however, Bob should be careful not to allow Eve to make a ``faked'' reference pulse which is accepted by Bob's classical detector but causes no clicks at his single-photon detector; using a local oscillator as proposed in Ref.\ \cite{koashi_2004} is a good solution to this problem; insecure implementations of B92 that do not use homodyne measurement have to be avoided \cite{buttler_1998,durt_1999_and_reply}). The modification of the BB84 protocol in Refs.\ \cite{lagasse_2005patent,nielsen_2001}, with a single detector randomly chosen via phase modulator setting to detect either a 0 or 1 bit, is not vulnerable for the same reason.\footnote{Although the B92 protocol and the modification of the BB84 protocol in Refs.\ \cite{lagasse_2005patent,nielsen_2001} are not affected by the attack described in the present paper, they are instead vulnerable to another attack. These protocols apply the key bit values directly at Bob's phase modulator, encoded in the phase shift settings. This makes them vulnerable to the large-pulse attack \cite{vakhitov_2001,gisin_2006}: The phase shift settings could be read by Eve from Bob's modulator using external light pulses which do not have to be very bright. The Scarani-Acin-Ribordy-Gisin 2004 (SARG04) protocol \cite{scarani_acin,acin_gisin,branciard_2005} also applies the key bit values at Bob's modulator. Other protocols only apply detection bases at Bob's modulator, which makes them less vulnerable to the large-pulse attack.}
The six-state protocol \cite{bruss_sixstates_orig,bechmann-pasquinucci_1999,bennett_1984_ibm} seems not to be vulnerable (though we note that a faked-states attack along the lines of Sec.~II on the six-state protocol gives 25\% QBER in the case of total efficiency mismatch, while the straight intercept-resend attack results in 33.3\% QBER).

On the other hand, the SARG04 protocol \cite{scarani_acin,acin_gisin,branciard_2005} is vulnerable to this attack. Also, faked states exploiting detector efficiency mismatch can be constructed for energy-time encoding and differential phase shift keying QKD schemes \cite{tittel_2000,nambu_2004,inoue_2002,buttler_2002,takesue_2005}; see examples of faked states in Ref.\ \cite{quant-ph-0702262}.

Implementations with a source of entangled pairs placed {\em outside} of Alice and Bob (as opposed to using it inside Alice to prepare the states) give Eve additional degrees of freedom to run this attack. When photons travel from Alice to Bob, Eve can completely block only one of Bob's bases (one detector is blocked by timing and the other by destructive interference in this basis). This allows to eavesdrop on the protocols that use two bases (BB84, SARG04), but not on the protocols that use three bases (six-state protocol, Ekert protocol \cite{ekert} if it is implemented with an entangled pair source inside Alice). However when photons travel from the entangled pair source to Alice and Bob with both paths accessible to Eve, she can replace the entangled pair source with a faked one, generating two faked states synchronously: one for Alice and one for Bob. She can generate a pair of faked states that block completely one basis at Alice and another basis at Bob. Then Alice and Bob only get coincidence clicks in the same basis when they choose the third basis in the protocol. This allows to eavesdrop on the six-state protocol \cite{bruss_sixstates_orig,bechmann-pasquinucci_1999} if it is implemented in an entangled pair version, with the source of entangled pairs placed between Alice and Bob. Also a set of faked states can be constructed for the Ekert protocol (at least if it is implemented as described in Ref.\ \cite{ekert} with no additional consistency checks besides checking that $S = - 2\sqrt{2}$ \cite{quant-ph-0702262}).

Throughout the paper, Eve used time $t$ as a control parameter to alter detector efficiencies. We note that $t$ could in principle be regarded as a general control parameter allowing Eve to change Bob's detector efficiencies. It could be not necessarily time but, e.g., polarization or wavelength. For instance, in up-conversion single-photon detectors \cite{thew_2006,langrock_2005,diamanti_2005} hardware gating of detectors is removed, but a narrow wavelength selectivity is introduced instead. Eve could try to use the wavelength of pulses instead of time to run this attack.

Finally we note that Qi {\em et al.} have recently proposed an interesting modification of our attack \cite{QuantInfComp-7-p73}.

\appendix*
\section{Quantum nondemolition measurement of qubit timing}

Here we will show that Eve can perform quantum nondemolition measurements of the timing of the qubits, and collapse Alice's photon pulses into arbitrarily narrow pulses. This measurement does not affect the degrees of freedom encoding the qubit. While (time-bin) phase-encoded qubits are considered here, one may treat other encodings in a similar way.

The phase-encoded qubit is denoted $|\varphi\rangle_{t_0}$. Here, $\varphi$ is the phase difference between the two pulses ($0\textdegree$, $90\textdegree$, $180\textdegree$, or $270\textdegree$), and $t_0$ is the (absolute) timing of the pulses, i.e., the time of the peak of the first pulse. If we assume that $|\varphi\rangle_{t_0}$ is a single-photon state,\footnote{Coherent pulses can be treated along the same lines.} it can be expressed as
\begin{equation}
\label{qubitrepr}
|\varphi\rangle_{t_0}=\frac{1}{\sqrt 2} \left(a^\dagger_{t_0}+e^{i\varphi}a^\dagger_{t_0+\tau}\right)|0\rangle,
\end{equation}
where $|0\rangle$ is the vacuum state of the single optical mode, $\tau$ is the time delay between the two pulses, and
\begin{equation}
\label{pulsecreation}
a_{t_0}^\dagger=\int dt\  \xi(t,t_0)a^\dagger(t).
\end{equation}
In Eq.\ \eqref{pulsecreation}, $a^\dagger(t)$ is the continuous-time creation operator \cite{loudon} of the optical mode. The operator satisfies the commutator relation $[a(t),a^\dagger(t')]=\delta(t-t')$. The function $\xi(t,t_0)$ represents, for instance, a Gaussian pulse shape:
\begin{equation}
\label{gaussfun}
\xi(t,t_0)=(2\Delta^2/\pi)^{1/4}\exp
\big[-i\omega_0 (t-t_0)-\Delta^2(t-t_0)^2\big].
\end{equation}
Here $\omega_0$ and $\Delta$ are the central frequency
and pulse bandwidth, respectively. The duration $t_{\Delta}$ of the pulse is of the order $1/\Delta$, and satisfies $t_{\Delta} \ll \tau$.

If Eve wants to measure the timing of a qubit pulse pair, she should do a nondemolition measurement that does not affect the degrees of freedom encoding the qubit. She divides the pulse time range $[t_0-t_{\Delta}/2, t_0+t_{\Delta}/2]$ into small intervals $T_i=[t_0-t_{\Delta}/2+i\Delta t,t_0-t_{\Delta}/2+(i+1)\Delta t]$, where $i$ is a positive integer and $\Delta t$ is her time resolution. [We assume that she has rough estimates of $t_0$ and $t_\Delta$ {\em a~priori}, with precision better than (of the order of) $t_{\Delta}$. Moreover, she knows $\tau$ with precision better than $\Delta t$.] The non-demolition measurement is described formally by the projectors
\begin{equation}
\label{projectors}
{P(T_i)=\int_{T_i} dt \left[a^\dagger(t)|0\rangle\langle 0|a(t)+a^\dagger(t+\tau)|0\rangle\langle 0|a(t+\tau)\right]}.
\end{equation}
Note that $P(T_i)P(T_j)=\delta_{ij}P(T_i)$ and $\sum_i P(T_i)=1$ in the Hilbert space spanned by the signal states \eqref{qubitrepr}, so this is a valid quantum mechanical projective measurement \cite{nielsen_chuang}. Moreover, when the projectors $P(T_i)$ act on the state \eqref{qubitrepr} the pulse width of each of the two pulses collapses to a smaller pulse width $\Delta t$; however the qubit encoding is not affected. In other words, Eve compresses the pulses and obtains the timing information $i$.

One way to implement this measurement is first to switch the two pulses into two optical modes $a$ and $b$. The first pulse is then delayed by $\tau$ so that the two pulses arrive at the measuring device simultaneously. The signal state \eqref{qubitrepr} can now be expressed as $|\varphi\rangle=\frac{1}{\sqrt 2}\left(a^\dagger+e^{i\varphi}b^\dagger\right)|00\rangle=\frac{1}{\sqrt 2}\left(|10\rangle+e^{i\varphi}|01\rangle\right)$, omitting the time notation for simplicity. Now, Eve lets a probe (a simple quantum computer) interact unitarily with the signal state, described as follows: $|00\rangle|0\rangle\rightarrow |00\rangle|0\rangle$, $|01\rangle|0\rangle\rightarrow |01\rangle|1\rangle$, $|10\rangle|0\rangle\rightarrow |10\rangle|1\rangle$. Here the last state in the product denotes that of the probe. Since $|00\rangle|0\rangle\rightarrow |00\rangle|0\rangle$ and $|\varphi\rangle|0\rangle\rightarrow |\varphi\rangle|1\rangle$, Eve will detect the presence of the qubit without disturbing it. Moreover, if her measurement device is sufficiently fast, she is able to obtain the timing (and the pulses will collapse into shorter ones).

\bibliography{qkdbib}

\end{document}